\documentclass[12pt]{article}
\def\be{\begin{equation}}
\def\ee{\end{equation}}
\def\bea{\begin{eqnarray}}
\def\eea{\end{eqnarray}}
\usepackage{graphicx}

\catcode`\@=11
\def\lsim{\mathrel{\mathpalette\@versim<}}
\def\gsim{\mathrel{\mathpalette\@versim>}}
\def\@versim#1#2{\vcenter{\offinterlineskip
\ialign{$\m@th#1\hfil##\hfil$\crcr#2\crcr\sim\crcr } }}
\catcode`\@=12

\catcode`@=12
\begin{document}

\thispagestyle{empty}
\begin{flushright}
UCRHEP-T511\\
December 2011\
\end{flushright}
\vspace{0.3in}
\begin{center}
{\Large \bf Scalar Neutrino as Asymmetric Dark Matter:\\ 
Radiative Neutrino Mass and Leptogenesis\\}
\vspace{1.2in}
{\bf Ernest Ma$^a$ and Utpal Sarkar$^{b,c}$\\}
\vspace{0.2in}
{$^a$ \sl Department of Physics and Astronomy, University of California,\\ 
Riverside, California 92521, USA\\}
{$^b$ \sl Physical Reserach Laboratory, Ahmedabad 380009, India\\}
{$^c$ \sl Physics Department and McDonnell Center for the Space Sciences, \\
Washington University, St. Louis, Missouri 63130, USA\\}
\end{center}
\vspace{1.2in}
\begin{abstract}\
In the Minimal Supersymmetric Standard Model (MSSM), the scalar neutrino 
$\tilde{\nu}_L$ has odd $R$ parity, yet it has long been eliminated as a 
dark-matter candidate because it scatters elastically off nuclei through 
the $Z$ boson, yielding a cross section many orders of magnitude above 
the experimental limit.  We show how it can be reinstated as a dark-matter 
candidate by splitting the masses of its real and imaginary parts in an 
extension of the MSSM with scalar triplets.  As a result, radiative Majorana 
neutrino masses are also generated.  In addition, decays of the scalar 
triplets relate the abundance of this asymmetric dark matter to the baryon 
asymmetry of the Universe through leptogenesis.
\end{abstract}

\newpage
\baselineskip 24pt

The imposition of $R$ parity, i.e. $R \equiv (-1)^{3B+L+2j}$, in the Minimal 
Supersymmetric Standard Model (MSSM) of particle interactions serves at 
least two purposes.  One is to avoid proton decay in its renormalizable 
interactions; the other is to establish a dark-matter candidate which 
is neutral and stable, i.e. odd under $R$.  This candidate particle may be 
a boson or a fermion.  If it is a boson, then it should be the lightest of 
three scalar neutrinos $\tilde{\nu}_L$.  If it is a fermion, then it should be 
the lightest of four neutralinos, i.e. the $U(1)$ and neutral $SU(2)$ 
gauginos and the two neutral higgsinos.  However, a scalar neutrino 
scatters elastically off nuclei with an amplitude mediated by the $Z$ 
boson, yielding a cross section many orders of magnitude above the present 
experimental limit, so it was eliminated as a dark-matter candidate many 
years ago.  As for the lightest neutralino, which is a linear combination 
of gauginos (which do not couple to the $Z$ boson) and higgsinos (which do), 
it is still considered as the canonical candidate for dark matter.

To reinstate the scalar neutrino as a dark-matter candidate, its elastic 
scattering with nuclei must be suppressed and this is easily achieved by 
splitting the mass of its real and imaginary components.  The reason is 
that the coupling of the vector $Z$ boson to $\tilde{\nu}_L = (\tilde{\nu}_1 
+ i \tilde{\nu}_2)/\sqrt{2}$ is of the form $Z \tilde{\nu}_1 \tilde{\nu}_2$, 
so if the mass gap is greater than about 100 keV, this process is forbidden 
by kinematics in the nuclear elastic recoil experiments.

There are now two issues to be considered. (1) How is this splitting 
achieved?  A mass splitting term $\tilde{\nu}_L \tilde{\nu}_L$ cannot be put 
in by hand, 
because it is not invariant under the $SU(2)_L \times U(1)_Y$ gauge 
symmetry of the MSSM.  If it is simply assumed to be an effective term 
without specifying its underlying origin, then it cannot be guaranteed 
that whatever conclusion is drawn from its existence will not be affected 
by the actual dynamics which generated it in the first place.  Here we 
assume that it comes from the gauge-invariant term $\Delta^0_1 \tilde{\nu}_L 
\tilde{\nu}_L - \sqrt{2} \Delta^+_1 \tilde{\nu}_L \tilde{e}_L - \Delta^{++}_1 
\tilde{e}_L \tilde{e}_L$, where $\Delta_1 = (\Delta^{++}_1, \Delta^+_1, 
\Delta^0_1)$ is a scalar triplet, with a vacuum expectation value 
$\langle \Delta^0_1 \rangle = u_1$. (2) Once the specific origin of this 
splitting is identified, what are its  physical consequences?  The first 
is of course neutrino mass.  Since $\tilde{\nu}_L$ carries lepton number $L$, 
the induced mass splitting term $\tilde{\nu}_L \tilde{\nu}_L$ breaks $L$ 
to $(-1)^L$.  
The observed neutrinos must then have Majorana masses and a 
radiative contribution must exist through the exchange of $\tilde{\nu}_L$ 
and neutralinos in one loop.  More importantly, the scalar triplet $\Delta_1$ 
should also couple to the neutrinos directly which then obtain masses through 
$u_1$ in the well-known manner of the Type II seesaw.  This latter 
would imply a very small $u_1$, much less than 100 keV, thus invalidating 
the interpretation of $\tilde{\nu}_1$ as dark matter.

In the following we forbid the dimension-four term $\Delta^0_1 \nu_L \nu_L$ by 
assigning $L=0$ to $\Delta_{1,2}$ where $\Delta_2 = (\Delta^{0}_2, \Delta^-_2, 
\Delta^{--}_2)$ and insisting that $L$ be conserved by all 
dimension-four terms of the supersymmetric Lagrangian of this model.  
We then break the supersymmetry by soft terms which are allowed to 
break $L$ to $(-1)^L$ as well, i.e. the dimension-three term $\Delta^0_1 
\tilde{\nu}_L \tilde{\nu}_L$.  We then show that neutrinos do acquire 
radiative Majorana masses~\cite{m06} in this case, but they are only compatible 
with $\tilde{\nu}_1$ as dark matter if the $U(1)$ and $SU(2)$ gaugino masses 
have opposite signs, a phenomenological possibility that has been largely 
overlooked.  We also show how the decays of $\Delta_{1,2}$ result~\cite{ms98} 
in both a lepton asymmetry and an asymmetry in $\tilde{\nu}_L$, with its 
relic density determined by the subsequent annihilation of $\tilde{\nu}_L 
\tilde{\nu}_L$ into $\nu_L \nu_L$.  Note that the mass splitting of the 
scalar neutrino is not induced by heavy singlet (right-handed) neutrino 
superfields through mixing.  If it were~\cite{gh97,hmm98}, then there would 
also be a tree-level neutrino mass from the Type I seesaw.  If the inverse 
seesaw mechanism were used instead~\cite{a08}, then again there would be both 
a tree-level mass and a loop-induced mass.  In our case, only the latter 
occurs and as we show later in Eq.~(3), this is a crucial condition for 
$\tilde{\nu}_1$ to be a viable dark-matter candidate.

The superpotential of this model is given by
\begin{eqnarray}
W &=& \mu \hat{\Phi}_1 \hat{\Phi}_2 + f^e_{ij} \hat{\Phi}_1 \hat{L}_i 
\hat{e}^c_j + f^d_{ij} \hat{\Phi}_1 \hat{Q}_i \hat{d}^c_j 
+ f^u_{ij} \hat{\Phi}_2 \hat{Q}_i \hat{u}^c_j \nonumber \\ &+& M 
\hat{\Delta}_1 \hat{\Delta}_2 + f_1 \hat{\Delta}_1 \hat{\Phi}_1 \hat{\Phi}_1 
+ f_2 \hat{\Delta}_2 \hat{\Phi}_2 \hat{\Phi}_2, 
\end{eqnarray}
where $\hat{\Phi}_1 \sim (1,2,-1/2)$, $\hat{\Phi}_2 \sim (1,2,1/2)$, 
$\hat{L} \sim (1,2,-1/2)$, $\hat{e}^c \sim (1,1,1)$, $\hat{Q} \sim 
(3,2,1/6)$, $\hat{d}^c \sim (3^*,1,1/3)$, $\hat{u}^c \sim (3^*,1,-2/3)$, 
as in the MSSM.  The Higgs triplet superfields are $\hat{\Delta}_1 \sim 
(1,3,1)$ and $\hat{\Delta}_2 \sim (1,3,-1)$, which have been assigned 
lepton number $L=0$, so that the terms $\hat{\Delta}_1 \hat{L}_i \hat{L}_j$ 
are forbidden.

We allow $L$ to be broken by soft terms which also break the supersymmetry 
of the model, but only by two units, i.e. $\Delta L = \pm 2$.  This would 
forbid the bilinear $\tilde{L}_i \Phi_2$ and trilinear 
$\tilde{L}_i \tilde{L_j} \tilde{e}^c_k$, $\tilde{L}_i \tilde{Q}_j 
\tilde{d}^c$ terms, but allow the trilinear $\Delta_1 \tilde{L}_i 
\tilde{L}_j$ terms.  This pattern is stable because it is maintained 
by the residual $Z_2$ symmetry $(-1)^L$.  Since $\Delta_{1,2}$ mix 
through the soft $B M \Delta_1 \Delta_2$ term, the two resulting mass 
eigenstates both decay into states of $L=2$ as well as $L=0$, i.e. 
$\Phi_{1,2} \Phi_{1,2}$.  Leptogenesis~\cite{ms98} is then possible. 
Details of this supersymmetric scenario has been worked out 
previously~\cite{hms01,cs07}.  

\begin{figure}[thb]
\centerline{\includegraphics[width=2.75in]{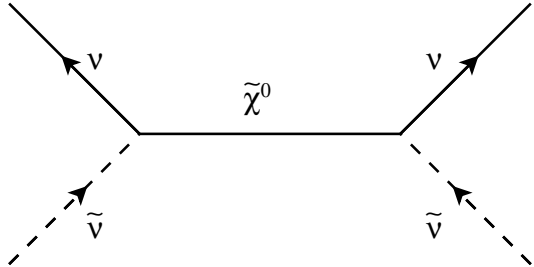}}
\caption{Annihilation of $\tilde \nu \tilde \nu \to \nu \nu$ via
neutralino exchange.}
\label{fig1}
\end{figure}

As a lepton asymmetry is established, there is also an asymmetry of the 
scalar neutrinos $\tilde{\nu}$.  This connection between visible and dark 
matter has been explored previously~\cite{hmw05,gs10,glsz11}.  It is also 
possible in the context of the radiative seesaw model of 
neutrino mass~\cite{m06} with the addition of heavy scalar triplets, as 
proposed recently~\cite{as11}.  As the Universe cools below $m_{\tilde{\nu}}$, 
the relic abundance of $\tilde{\nu}$ is determined by its annihilation 
cross sections with itself and with $\tilde{\nu}^*$.  The latter is very 
large, which means it contributes very little to the $\tilde{\nu}$ relic 
abundance.  The former is rather small, so its would-be relic density 
may be very large, but it is diminished by the $\tilde{\nu} - \tilde{\nu}^*$ 
asymmetry created earlier, so it may be just right. In  that case, it would 
be a good candidate for the dark matter of the Universe.  The $\tilde{\nu} 
\tilde{\nu}$ annihilation proceeds through neutralino exchange, as shown in 
Fig.~1.  In the $4 \times 4$ neutralino mass matrix, if the higgsino mass 
parameter $\mu$ is large, then the $2 \times 2$ gaugino mass matrix does 
not mix significantly with the $2 \times 2$ higgsino mass matrix, resulting 
in approximate mass eignevalues $m_{1,2}$ for the $U(1)$ and $SU(2)$ gauginos. 
In that case, this cross section $\times$ relative velocity is given by
\begin{equation}
\langle \sigma v \rangle = {g^4 \over 128 \pi c^4 m^2_{\tilde{\nu}}} \left( 
{s^2 y_1 \over y_1^2 + 1} + {c^2 y_2 \over y_2^2 + 1} \right)^2,
\end{equation}
where $s = \sin \theta_W$, $c = \cos \theta_W$, and 
$y_{1,2} = m_{1,2}/m_{\tilde{\nu}}$. 

\begin{figure}[htb]
\centerline{\includegraphics[width=3in]{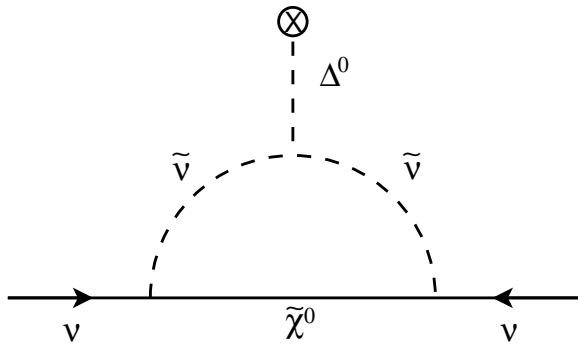}}
\caption{One-loop radiative Majorana neutrino mass via neutralino exchange. 
Lepton number $L$ becomes $(-1)^L$ through the soft 
$\Delta^0 \tilde{\nu} \tilde{\nu}$ term.}
\label{fig2}
\end{figure}

\begin{figure}[htb]
\centerline{\includegraphics[width=5.5in]{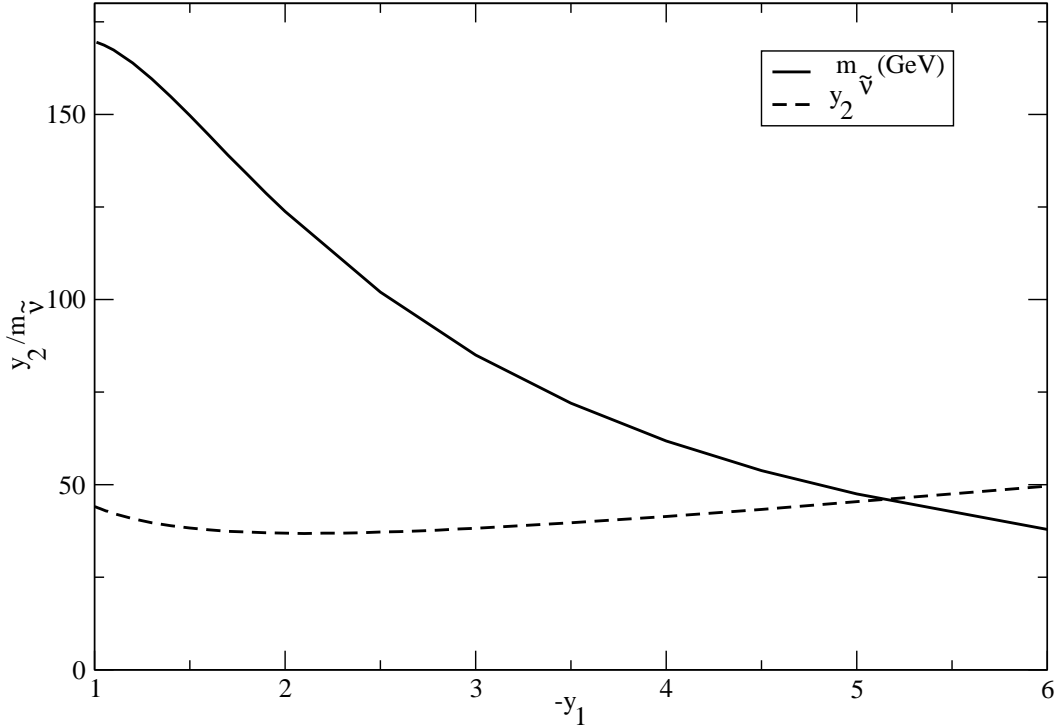}}
\vskip .15in
\caption{$y_2$ and $m_{\tilde \nu}$ are plotted against $-y_1$,
which correspond to $m_\nu/\Delta m_{\tilde{\nu}} = 0$ and 
$\langle \sigma v \rangle = 0.1$~pb.}
\label{fig3}
\end{figure}
For each $\tilde{\nu}$, a radiative neutrino mass is also generated in one loop 
by neutralino exchange, as shown in Fig.~2, resulting in~\cite{m06,gh97} 
\begin{equation}
{m_\nu \over \Delta m_{\tilde{\nu}}} = {g^2 \over 32 \pi^2 c^2} 
\left[ {s^2 y_1 \over y_1^2 - 1} \left( 1 - {y^2_1 \over y^2_1 - 1} \ln y^2_1 
\right) + {c^2 y_2 \over y^2_2 - 1} \left( 1 - {y^2_2 \over y^2_2 - 1} \ln y^2_2 
\right) \right].
\end{equation}
Since $\Delta m_{\tilde{\nu}} > 100$ keV is needed to suppress the interaction 
of $\tilde{\nu}_1$ with nuclei in underground direct-search experiments, 
and $m_\nu < 1$ eV for neutrino mass, the above ratio should be less than 
$10^{-5}$.  We assume that $|y_2| > |y_1| > 1$ and plot $y_2$ as a function 
of $-y_1$ so that $m_\nu/\Delta m_{\tilde{\nu}} = 0$ in Fig.~3. 
We also use Eq.~(2) to plot $m_{\tilde{\nu}}$ as a function of $-y_1$ for the 
particular value of $\langle \sigma v \rangle = 0.1$ pb.  This value is 
an order-of-magnitude smaller than the benchmark value of 1 pb for usual 
dark matter, and would overclose the Universe in that case.  Here, because 
of the $\tilde{\nu}$ asymmetry, the resulting relic density will be reduced 
and possible agreement with observation may be obtained.  Details will 
depend on other unknown parameters.  We simply use 0.1 pb to indicate 
the possible range of $m_{\tilde{\nu}}$ values in this scenario. 
We note that $m_{\tilde{\nu}} < m_Z/2$ is ruled out experimentally, 
because $Z \to \tilde{\nu}_1 \tilde{\nu}_2$ would then contribute to its 
invisible width, which already agrees very well with what is expected 
from the three known neutrinos of the Standard Model.

\begin{figure}[htb]
\centerline{\includegraphics[width=5.5in]{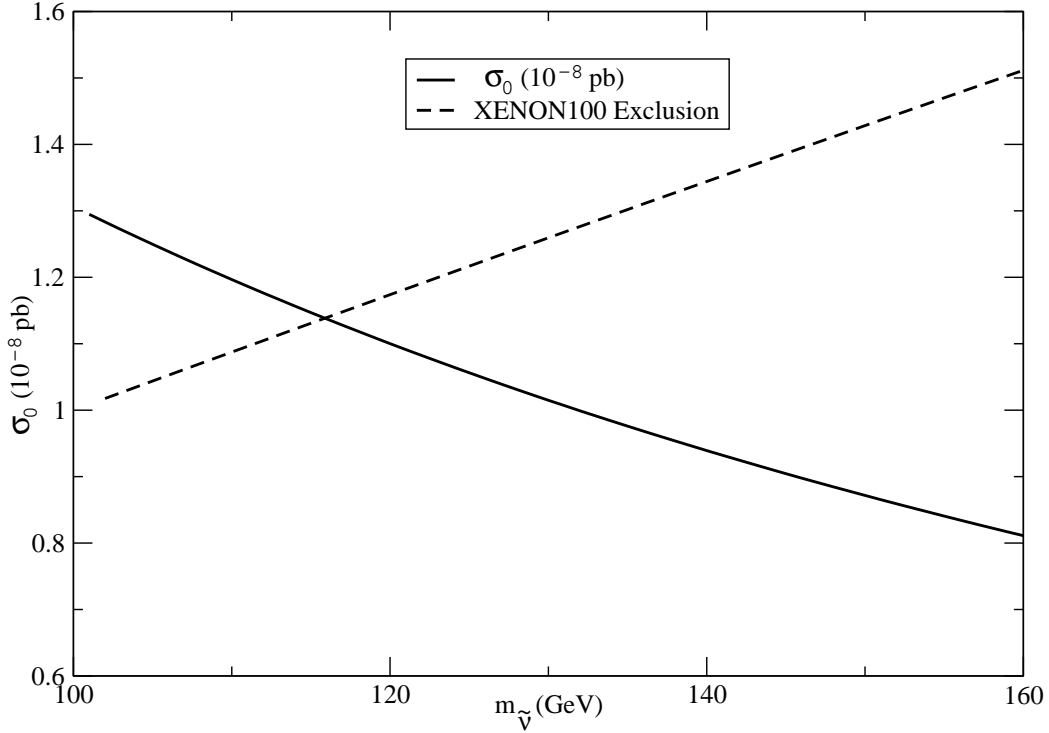}}
\vskip .15in
\caption{Spin-independent elastic scattering cross section of $\tilde{\nu}_1$ 
with $^{131}Xe$ through Higgs exchange is plotted together with the present 
experimental bound from the direct-search experiment XENON100, as a function of 
$m_{\tilde{\nu}}$.}
\label{fig4}
\end{figure}

In underground dark-matter direct-search experiments, the spin-independent 
elastic cross section for $\tilde{\nu}_1$ scattering off a nucleus of $Z$ 
protons and $A-Z$ neutrons normalized to one nucleon is given by
\begin{equation}
\sigma_0 = {1 \over \pi} \left( {m_N \over m_{\tilde{\nu}} + A m_N} \right)^2 
\left| {Z f_p + (A-Z) f_n \over A} \right|^2,
\end{equation}
where $m_N$ is the mass of a nucleon, and $f_{p,n}$ come from Higgs 
exchange~\cite{hiny11}:
\begin{eqnarray}
{f_p \over m_p} &=& \left( - {0.075 \over 4} - {0.925(3.51) \over 54} \right) 
{g^2 \over c^2 m^2_\phi}, \\ 
{f_n \over m_n} &=& \left( - {0.078 \over 4} - {0.922(3.51) \over 54} \right) 
{g^2 \over c^2 m^2_\phi}.
\end{eqnarray}
Assuming an effective $m_\phi = 130$ GeV and using $Z=54$ and $A-Z=77$ for 
$^{131}Xe$, we plot $\sigma_0$ as a function of $m_{\tilde{\nu}}$ in Fig.~4.  
Note for $m_{\tilde{\nu}}=120$ GeV, $\sigma_0 = 1.1 \times 10^{-8}$ pb, 
which is just below the upper limit of the 2011 XENON100 
exclusion~\cite{xenon11}.  The allowed range for $m_{\tilde{\nu}}$ is thus 
above 116 GeV and probably no greater than about 170 GeV from Fig.~3.

If $\Delta m_{\tilde{\nu}} < 2 m_e$, the decay of $\tilde{\nu}_2$ is only to 
$\tilde{\nu}_1 \nu \bar{\nu}$ through $Z$ exchange.  This decay width is 
given by
\begin{equation}
\Gamma (\tilde{\nu}_2 \to \tilde{\nu}_1  \nu \bar{\nu}) = 
{G_F^2 (\Delta m_{\tilde{\nu}})^5 \over 60 \pi^3}.
\end{equation}
For $100~{\rm keV} < \Delta m_{\tilde{\nu}} < 1~{\rm MeV}$, the corresponding 
lifetime is of order $10^9$ to $10^4$ seconds.  This means that in the early 
Universe, $\tilde{\nu}_L \tilde{\nu}_L$ annihilation should be considered for 
determining the relic abundance of $\tilde{\nu}_L$, but at present 
only $\tilde{\nu}_1$ survives as dark matter.  If a concentration of 
$\tilde{\nu}_1$ has accumulated inside the sun or the earth, $\tilde{\nu}_1 
\tilde{\nu}_1$ annihilation to two monoenergetic neutrinos~\cite{f11} 
would be a spectacular indication of this scenario.

If $\tilde{\nu}_1$ is dark matter, its production at the Large Hadron Collider 
(LHC) must always be accompanied by a lepton.  If it comes from the decay of 
the $U(1)$ gaugino, i.e. $\tilde{B} \to \bar{\nu} \tilde{\nu}$, then it 
is completely invisible.  If it comes from a chargino, i.e. $\tilde{\chi}^+ \to 
e^+ \tilde{\nu}$, then it may be discovered through its missing energy and 
large mass.  More detailed study of this scenario is required.

The Higgs triplet scalars $\Delta_{1,2}$ are the common origins of both 
the observed baryon asymmetry and the asymmetric $\tilde{\nu}_L$ dark matter 
of the Universe.  They are likely to be very heavy, say of order 
$10^{8}$ GeV, in which case they are not accessible at the LHC.  On the 
other hand, resonant leptogenesis may occur naturally in this 
scenario~\cite{cs07}, which would allow them to be at the 
TeV scale.  In that case, the direct decay $\Delta_1^{++} \to \tilde{e}^+_i 
\tilde{e}^+_j$ would serve to map out the neutrino mass matrix, in analogy 
to the previously proposed simple scenario~\cite{mrs00,mrs01}, where 
$\Delta^{++} \to e^+_i e^+_j$.

In conclusion, we have proposed that the dark matter of the Universe is 
the real component of the lightest scalar neutrino $\tilde{\nu}_1$ in the 
Minimal Supersymmetric Standard Model.  To implement this unconventional 
scenario, we add Higgs triplet superfields $\hat{\Delta}_{1,2}$, so that 
the observed neutrinos acquire small radiative Majorana masses from the 
mass splitting terms $\Delta_1 \tilde{\nu}_L \tilde{\nu}_L$.  The fact that 
$m_\nu/\Delta m_{\tilde{\nu}} < 10^{-5}$ forces the $U(1)$ and $SU(2)$ gaugino 
masses to have opposite signs.  Using the canonical 1 pb cross section 
for dark matter and the latest XENON100 data, the allowed mass range of 
$\tilde{\nu}_1$ is above 116 GeV and probably no greater than about 170 GeV, 
assuming a Higgs-boson mass of 130 GeV.

This research is supported in part by the U.~S.~Department of Energy 
under Grant No.~DE-AC02-06CH11357. One of us (US) thanks R. Cowsik for
arranging his visit as the Clark Way Harrison Visiting Professor at 
Washington University..

\baselineskip 16pt
\bibliographystyle{unsrt}

\end{document}